\documentclass[twocolumn,showpacs,floatfix,aps]{revtex4}
\usepackage{amssymb,amsmath}
\usepackage[dvips]{graphics}
\usepackage{epsfig}
\newcommand{\bea}{\begin{array}}
\newcommand{\ear}{\end{array}}

\newcommand{\bege}{\begin{equation}}
\newcommand{\enge}{\end{equation}}

\newcommand{\beq}{\begin{eqnarray}}\newcommand{\benu}{\begin{enumerate}}\newcommand{\enu}{\end{enumerate}}
\newcommand{\eeq}{\end{eqnarray}}

\newcommand{\lie}{\text{{\it\char'44}}}

\begin{document}

\title{Could the variation in quasar luminosity, due to extra dimension 3-brane in RS model, be measurable?}

\author{R. da Rocha}
\email{roldao@ifi.unicamp.br}
\affiliation{IFGW, Universidade Estadual de Campinas,\\
CP 6165, 13083-970 Campinas, SP, Brazil.}

\author{C. H. Coimbra-Ara\'ujo}
\email{carlos@astro.iag.usp.br}
\affiliation{Departamento de Astronomia, Universidade de S\~ao Paulo, 05508-900
S\~ao Paulo, SP, Brazil.\\and\\
IFGW, Universidade Estadual de Campinas,\\
CP 6165, 13083-970 Campinas, SP, Brazil.}


\begin{abstract}
We propose an alternative theoretical approach  showing how the existence of an extra dimension in RS model can estimate 
the correction in the Schwarzschild radius of black holes, and consequently its measurability in terms of the variation of quasar luminosity,
which can be caused by a imprint of an extra dimension endowing 
the geometry of a brane-world scenario in an AdS$_5$ bulk.
This paper is intended to investigate the variation of luminosity due to accretion of gas in black holes (BHs)
in the center of quasars, besides also investigating the variation of luminosity in supermassive BHs by brane-world effects, 
using RS model. 
 
\end{abstract}

\maketitle

\section{Introduction}

The possibility concerning the existence of extra dimensions is one of the most astonishing 
aspects on string theory and the formalism of $p$-branes.
In spite of this  possibility, extra dimensions still remain up to now unaccessible and obliterated to experiments.
An alternative approach to the compactification of extra dimensions, provided by, e.g., Kaluza-Klein (KK) and string theories \cite{gr,zwi,zwi1,zwi2}, 
involves an extra dimension which is not compactified, as pointed by, e.g.,  RS model \cite{Randall1,Randall2}. 
This extra dimension implies
deviations on Newton's law of gravity at scales below about 0.1 mm, where objects may be indeed gravitating in more
dimensions. The electromagnetic, weak and strong forces, as well as all the matter
in the universe, would be trapped on a brane with three spatial dimensions, and only gravitons would be allowed to leave the surface
 and move into the full bulk, constituted by an AdS$_5$ spacetime, as prescribed by, e.g., 
 in RS model \cite{Randall1,Randall2}.

At low energies, gravity is localized on the brane and general relativity is recovered, but at high energies,
 significant changes are introduced in gravitational dynamics, forcing general relativity to break down
to be overcome by a quantum gravity theory \cite{rov}.
A plausible reason for the gravitational force appear to be so weak in relation to other forces can be its dilution in possibly existing extra dimensions
related to a bulk, where $p$-branes \cite{gr,zwi, zwi1, zwi2, Townsend} are embedded. $p$-branes 
are good candidates for brane-worlds \cite{ken} because they possess gauge symmetries \cite{zwi, zwi1, zwi2} and automatically incorporate a quantum theory of gravity. The
gauge symmetry arises from open strings, which can collide to form a closed string that can leak into the higher-dimensional
bulk. The simplest excitation modes of these closed strings
correspond precisely to gravitons.
An alternative scenario can be achieved by Randall-Sundrum 
  model (RS) \cite{Randall1,Randall2}, 
which induces a volcano barrier-shaped effective potential for gravitons around the brane \cite{Likken}.
The corresponding spectrum of gravitational perturbations has a massless bound state on the brane,
and a continuum of bulk modes with suppressed couplings to brane fields. These bulk modes introduce small
corrections at short distances, and the introduction of more compact dimensions does not affect the localization of matter fields.
However, true localization takes place only for massless fields \cite{Gregory}, and in the massive case the bound state becomes metastable,
being able to leak into the extra space. This is shown to be exactly the case for astrophysical massive objects,
where highly energetic stars and the process of gravitational collapse, which can originate black holes, leads to deviations from the $4D$ general relativity problem.
There are other interesting and astonishing features concerning RS models, such as the AdS/CFT correspondence of a RS infinite AdS$_5$ brane-world, without matter 
fields on the brane, and 4-dimensional general relativity coupled to conformal fields \cite{Randall1,Randall2,Maartens}. 

We precisely investigate the consequences of the deviation of a Schwarzschild-like term in a 5$D$ spacetime metric, predicted 
by RS1 model in the correction of the Schwarzschild radius of a BH. We show that, for fixed effective extra dimension size, supermassive BHs (SMBHs)  
give the upper limit of variation in luminosity of quasars, and although the method used holds for any other kind of BH, such as mini-BHs and 
stellar-mass ones, we shall use SMBHs parameters, where the effects are seen to be more notorious.
 It is also analyzed how the quasar luminosity variation behaves as a function of the AdS$_5$ bulk radius 
$\ell$, for various values of BH masses, from $10$ to $10^6$ solar masses. 

The search for observational evidence of higher-dimensional gravity is an
important way to test the ideas that have being come from string theory. This
evidence could be observed in particle accelerators or gravitational wave
detectors. The wave-form of gravitational waves produced by black holes, for
example, could carry an observational signature of extra dimensions, because
brane-world models introduce small corrections to the field equations at high
energies. But the observation of gravitational waves faces severe limitations
in the technological precision required for detection. This is a undeniable fact.
Possibly, an easier manner of testing extra dimensions can be via the
observation of signatures in the luminous spectrum of quasars and microquasars. This is the
goal of this paper, which is the first of a series of papers we shall present. Here we show the possibility of detecting brane-world
corrections for big quasars by their luminosity observation. In the next article we shall see that 
these corrections are more notorious in mini-BHs, where the Schwarzschild radius in a brane-world scenario 
shall be shown to be $10^4$ times bigger than standard Schwarzschild radius associated with mini-BHs. 
Indeed, mini-BHs are to be shown to be much more  
sensitive to brane-world effects.
In the last article of this series we also present an alternative 
possibility to detect electromagnetic KK modes due to perturbations in black strings \cite{ma,soda}.

This article  is organized as follows: in Section 2 after presenting Einstein equations
in AdS$_5$ bulk and discussing the relationship between the electric part of Weyl tensor 
and KK modes in RS1 model, the deviation in Newton's 4$D$ gravitational potential
is introduced in order to predict the deviation in Schwarzschild form and 
its consequences on the variation in quasar luminosity. For a static spherical metric 
on the brane the propagating effect of 5$D$ gravity is shown to 
arise only in the fourth order expansion in terms of the Talor's of the normal coordinate out of the brane. 
In Section 3 the variation in quasar luminosity is carefully investigated, by finding 
the correction in the Schwarzschild radius caused by brane-world effects. All results are illustrated by graphics 
and figures.

\section{Black holes on the brane}
In a brane-world scenario given by a 3-brane embedded in an AdS$_5$ bulk the Einstein field equations read 
\begin{eqnarray}\label{123}
&&G_{\mu\nu} = -\frac{1}{2}{\Lambda}_5g_{\mu\nu}\nonumber\\
&&+ \frac{1}{4}\kappa_5^4\left[TT_{\mu\nu} - T^\alpha _{ \nu}T_{\mu \alpha} + \frac{1}{2}g_{\mu\nu}(T^2 - T_{\alpha\beta}^{\;\;\;\;\alpha\beta})\right] - E_{\mu\nu},\nonumber
\end{eqnarray}
\noindent where $T = T_\alpha^{\;\;\alpha}$ denotes the trace of the momentum-energy tensor $T_{\mu\nu}$, $\Lambda_5$ denotes 
the 5-dimensional cosmological AdS$_5$ bulk constant, and  $E_{\mu\nu}$ denotes the `electric' components of the Weyl tensor, that can be expressed by means of the
extrinsic curvature components $K_{\mu\nu} = -\displaystyle \frac{1}{2} \lie_n g_{\mu\nu}$ by \cite{soda}
\begin{equation}
E_{\mu\nu} = \lie_n K_{\mu\nu} + K_{\mu}^{\;\;\alpha}K_{\alpha\nu} - \frac{1}{\ell^2}g_{\mu\nu}
\end{equation}\noindent where $\ell$ denotes the AdS$_5$ bulk curvature radius. It corresponds equivalently 
to the effective size of the extra dimension probed by a 5$D$ graviton \cite{Likken, Randall1,Randall2,Maartens}
The constant  $\kappa_5 = 8\pi G_5$, where 
$G_5$ denotes the 5-dimensional Newton gravitational constant, that can be related to the
4-dimensional gravitational constant $G$ by $G_5 = G\ell_{\rm Planck}$, where $\ell_{\rm Planck} = \sqrt{G\hbar/c^3}$ is the Planck length.

As indicated in \cite{Randall1,Maartens}, ``table-top tests of Newton's law currently find no deviations down to the order
of 0.1 mm'', so that $\ell \lesssim $ 0.1 mm. Empar\'an et al \cite{emparan} provides a more accurate magnitude limit improvement on the AdS$_5$ curvature $\ell$, 
by analyzing the existence of stellar-mass BHs
on long time scales and of BH X-ray binaries. In this paper we relax the stringency $\ell \lesssim 0.01$ mm to the former table-top limit
$\ell \lesssim $ 0.1 mm. 

The Weyl `electric' term $E_{\mu\nu}$  carries an imprint of high-energy effects sourcing KK modes. It means that highly energetic
stars and the process of gravitational collapse, and naturally  BHs, lead to deviations from the 4-dimensional general
relativity problem. This occurs basically because the gravitational collapse unavoidably produces energies high enough to make
these corrections significant. From the brane-observer viewpoint, the KK corrections in $E_{\mu\nu}$ are nonlocal, since they
incorporate 5-dimensional gravity wave modes. These nonlocal corrections cannot be determined purely from data on the brane \cite{Maartens}.
The component $E_{\mu\nu}$ also carries information about the collapse process of BHs.
In the perturbative analysis of Randall-Sundrum (RS) positive tension 3-brane, KK modes consist of a continuous spectrum without any gap. It
generates a correction in the gravitational potential $V(r) =\frac{GM}{c^2r}$ to 4$D$ gravity at low energies from extra-dimensional effects \cite{Maartens}, 
which is
given by \cite{Randall1,Randall2}
\begin{equation}\label{potential}
V(r) = \frac{GM}{c^2r}\left[1 + \frac{2\ell^2}{3r^2} + \mathcal{O}\left(\frac{\ell}{r}\right)^4\right].
\end{equation}
\noindent
The KK modes that generate this correction are responsible for a nonzero $E_{\mu\nu}$. This term carries the modification to the weak-field field equations, as we have
already seen.
 The Gaussian coordinate $y$ denotes hereon the direction normal out of the brane into the AdS$_5$ bulk, in each point of the 3-brane\footnote{In general 
the vector field cannot be globally defined on the brane, and it is only possible if the 3-brane is considered to be parallelizable.}.

\begin{figure}
\includegraphics[width=8.7cm]{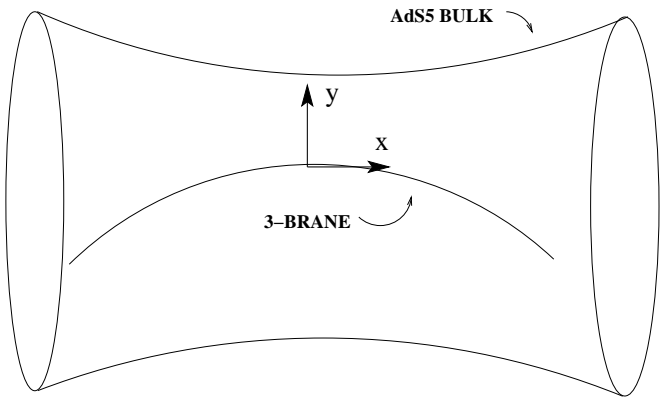}
  \caption{\small Schematic diagram of a slice of a 3-brane embedded in an AdS$_5$ bulk. The Gaussian coordinate
$y$ is normal to the brane and $x$ denotes spacetime coordinates in the brane.}
\label{fig:ads5}
\end{figure}

The RS metric is in general expressed as
\begin{equation}
^{(5)}ds^2 = e^{-2k|y|}g_{\mu\nu}dx^{\mu}dx^{\nu} + dy^2,
\end{equation}
\noindent
where $k^2 = 3/(2\ell^2)$,  
and the term $e^{-2k|y|}$ is called \emph{the warp factor} \cite{Randall1,Randall2,Maartens}, which
reflects the confinement role of the bulk cosmological constant $\Lambda_5$, preventing gravity from leaking
into the extra dimension at low energies \cite{Maartens,Randall1,Randall2}. 
The term $|y|$ clearly provides the $\mathbb{Z}_2$ symmetry
of the 3-brane at $y=0$. 

Concerning the anti-de Sitter (AdS$_5$) bulk, the cosmological constant can be written as $\Lambda_5 = -6/\ell^2$ and the brane is localized
at $y = 0$, where the metric recovers the usual aspect. 
The contribution of the bulk on the brane can be shown only to be due to the Einstein tensor, and can be expressed as
$\nabla_\nu G^{\mu\nu} = 0$, which implies that $\nabla_\nu (E^{\mu\nu} - S^{\mu\nu}) = 0$ \cite{Shiromizu},
where
\begin{equation}
S_{\mu\nu} := \frac{1}{4}\kappa_5^4\left[TT_{\mu\nu} - T^{\;\alpha} _{ \nu}T_{\mu \alpha} + \frac{1}{2}g_{\mu\nu}(T^2 - T_{\alpha\beta}^{\;\;\;\;\alpha\beta})\right]
\end{equation}
A vacuum on the brane, where $T_{\mu\nu} = 0$ outside a BH, implies that
\begin{equation}\label{21}
\nabla_\nu E^{\mu\nu} = 0.
\end{equation}
\noindent
Eqs.(\ref{21}) are referred to the nonlocal conservation equations. Other useful equations for the BH case are
\begin{equation}\label{ricci2}
G_{\mu\nu} = - \frac{1}{2}\Lambda_ 5g_{\mu\nu} - E_{\mu\nu}, \quad R = R^{\mu}_{\;\; \mu} = 0 = E^{\mu}_{\;\; \mu}.
\end{equation}
\noindent
 Therefore, a particular manner to express the vacuum field equations in the brane given by eq.(\ref{ricci2}) is
$E_{\mu\nu} = - R_{\mu\nu},$ 
where the bulk cosmological constant is incorporated to the warp factor in the metric.
One can use a Taylor expansion in order to probe properties of a static BH on the brane \cite{Da}, and for a vacuum brane metric,
we have, up to terms of order ${\mathcal O}(y^5)$ on, the following:
\begin{eqnarray}\label{metrica}
&&\negthickspace g_{\mu\nu}(x,y) = g_{\mu\nu}(x,0) - E_{\mu\nu}(x,0)y^2 - \frac{2}{\ell}E_{\mu\nu}(x,0)|y|^3\nonumber\\
&&\negthickspace+\frac{1}{12}\left[\left({\Box} - \frac{32}{\ell^2}\right)E_{\mu\nu} + 2R_{\mu\alpha\nu\beta}E^{\alpha\beta} + 6E_{\mu}^{\; \alpha}E_{\alpha\nu}\right]_{y=0}\negthickspace y^4
\nonumber\end{eqnarray}
\noindent where $\Box$ denotes the usual d'Alembertian.
It shows in particular that the propagating effect of $5D$ gravity arises only at the fourth order of the expansion. For a static spherical metric on the brane
given by \begin{equation}\label{124}
g_{\mu\nu}dx^{\mu}dx^{\nu} = - F(r)dt^2 + \frac{dr^2}{H(r)} + r^2d\Omega^2,
\end{equation}
\noindent where $d\Omega^2$ denotes the spherical 3-volume element related to the geometry of the 3-brane,
 the projected eletric component Weyl term on the brane is given by the expressions
\begin{eqnarray}
E_{00} &=& \frac{F}{r}\left(H' - \frac{1 - H}{r}\right),\;
E_{rr} = -\frac{1}{rH}\left(\frac{F'}{F} - \frac{1 - H}{r}\right),\nonumber\\
 E_{\theta\theta} &=& -1 + H +\frac{r}{2}H\left(\frac{F'}{F} + \frac{H'}{H}\right).
\end{eqnarray}
\noindent Note that in eq.(\ref{124}) the metric is led to the Schwarzschild one, if $F(r)$ equals $H(r)$.
The exact determination of these radial functions remains an open problem in BH theory on the brane \cite{Maartens,
rs05,rs06,rs07,rs08,rs09}.

These components allow one to evaluate the metric coefficients in eq.(\ref{metrica}). The area of the
$5D$ horizon is determined by $g_{\theta\theta}$. Defining $\psi(r)$ as the deviation from a Schwarzschild form for $H(r)$ \cite{Maartens,rs05,rs06,rs07,rs01,rs02,rs03,Gian}
\begin{equation}\label{h}
H(r) = 1 - \frac{2GM}{c^2r} + \psi(r),
\end{equation}
\noindent
where $M$ is constant, yields
\begin{eqnarray}\label{gtheta}
g_{\theta\theta}(r,y) &=& r^2  - \psi'\left(1 + \frac{2}{\ell}|y|\right)y^2\nonumber\\ +\negthickspace \negthickspace &&\negthickspace\negthickspace\left[\psi' + \frac{1}{2}(1 + \psi')(r\psi' - \psi)'\right]\frac{y^4}{6r^2} + \cdots
\end{eqnarray}
\noindent
It can be shown $\psi$ and its derivatives determine the change in the area of the horizon along the extra dimension \cite{Maartens}.
 For a large BH, with horizon scale $r \gg \ell$, it follows from eq.(\ref{potential}) that
\begin{equation}\label{psi}
\psi(r) \approx -\frac{4GM\ell^2}{3c^2r^3}.
\end{equation}
\noindent

\section{Variation in the luminosity of quasars and AdS curvature radius}
The observation of quasars (QSOs) in X-ray band can constrain the measure of the AdS$_5$ bulk curvature radius $\ell$, and indicate
 how the bulk is curled, from its geometrical and topological features.
QSOs are astrophysical objects that can be found at large astronomical distances (redshifts $z > 1$).
For a \emph{gedanken} experiment involving a static BH being accreted, in a simple model, the accretion eficiency $\eta$ is given by
\begin{equation}\label{eta}
\eta = \frac{GM}{6c^2R_{{\rm Sbrane}}},
\end{equation}
\noindent
where $R_{{\rm Sbrane}}$ is the Schwarzschild radius corrected for the case of brane-world effects. 
The luminosity $L$ due to accretion in a BH, that generates a quasar,
 is given by
\begin{equation}\label{dl}
L(\ell) = \eta(\ell) \dot{M}c^2,
\end{equation}
\noindent
where $\dot{M}$ denotes the accretion rate and depends on some specific model of accretion.

In order to estimate $R_{{\rm Sbrane}}$, fix $H(r) = 0$ in  eq.(\ref{h}), resulting in
\begin{equation}
1 - \frac{2GM}{c^2R_{{\rm Sbrane}}} - \frac{4GM\ell^2}{3c^2R_{{\rm Sbrane}}^3} = 0.
\end{equation}
\noindent This equation can be rewritten as 
\begin{equation}
R_{{\rm Sbrane}}^3 - \frac{2GM}{c^2}R_{{\rm Sbrane}}^2 - \frac{4GM\ell^2}{3c^2} = 0.
\end{equation}
\noindent Using Cardano's formul\ae\,\cite{card}, it follows that
\begin{equation}\label{111}
R_{{\rm Sbrane}} = (a + \sqrt{b})^{1/3} + (a - \sqrt{b})^{1/3} + \frac{2GM}{3c^2},\end{equation}
\noindent where 
\begin{eqnarray}
a &=& \frac{2GM}{3c^2}\left(\ell^2 + \frac{4G^2M^2}{9c^4}\right),\\\label{rs3}
 b&=&\frac{4G^2M^2\ell^2}{9c^4}\left(\ell^2 + \frac{8G^2M^2}{9c^4}\right).\label{rs4}
\end{eqnarray}
\noindent Writing $a$ and $b$ explicitly in terms of the Schwarzschild radius $R_S$ it follows from eqs.(\ref{rs3},\ref{rs4}) that
\begin{eqnarray}a&=& \frac{R_S}{3}\left(\ell^2 + \frac{R_S^2}{9}\right),\\\label{rs1}
b&=& \frac{R_S^2\ell^2}{9}\left(\ell^2 + \frac{2R_S^2}{9}\right).\label{rs2}
\end{eqnarray}
Now, substituting the values of $G$ and $c$ in the SI, and adopting $\ell \sim 0.1\, {\rm mm}$ and $M \sim 10^9 M_\odot$ (where $M_\odot \approx 2 \times 10^{33}$ g) 
denotes solar mass, corresponding to 
the mass of a SMBH, it follows from eq.(\ref{111})  that the correction in the Schwarzschild radius of a SMBH 
by brane-world effects is given by 
\begin{equation}\label{coor}
R_{{\rm Sbrane}} - R_S \sim 100\,{\rm m},\end{equation}
\noindent and since the Schwarzschild radius $R_S$ is defined as $\frac{2GM}{c^2} = 2.964444 \times 10^{12}\,{\rm m}$,  the relative error 
concerning the brane-world corrections in the Schwarzschild radius of a SMBH is given by 
\begin{equation}\label{razao}
1- \frac{R_S}{R_{{\rm Sbrane}}} \sim 10^{-10}\end{equation}
\noindent
These calculations shows that there exists a correction in the Schwarzschild radius of a SMBH caused by brane-world effects, although 
it is negligible. This tiny correction can be explained by the fact the event horizon of the SMBH is $10^{15}$ times bigger than the 
AdS$_5$ bulk curvature radius $\ell$. As shall be seen in a sequel paper these corrections 
are shown to be outstandingly wide in the case of mini-BHs, wherein the event horizon can be a lot of magnitude orders smaller than $\ell$.
As proved in \cite{rcp}, the solution above for $R_{{\rm Sbrane}}$ can be also found in terms of the curvature radius $\ell$.
It is then possible to find an expression for the luminosity $L$  in terms of the radius of curvature, regarding formul\ae\, (\ref{dl}).

Here we shall adopt the model of the accretion rate given by a disc accretion, given by \cite{shapiro} 
Having observational values for the luminosity $L$, it is  possible to estimate a value for $\ell$, given a BH accretion model.
For a tipical supermassive BH of $10^9 M_{\odot}$ in a massive quasar the accretion rate 
is given by 
\begin{equation}
\dot{M} \approx 2.1 \times 10^{16} {\rm kg}\, {\rm s}^{-1}
\end{equation}

Supposing the quasar radiates in Eddington limit, given by (see, e.g., \cite{shapiro})
\begin{equation}
L(\ell) = L_{{\rm Edd}} = 1.263 \times 10^{45}\left(\frac{M}{10^7 M_{\odot}}\right)\; {\rm erg\,  s^{-1}}
\end{equation}
\noindent
for a quasar with a supermassive BH of $10^9 M_{\odot}$, the luminosity is given by $L \sim 10^{47}\, {\rm erg\, s^{-1}}$.
From eqs.(\ref{eta}) and (\ref{dl}) the variation in quasar luminosity of a SMBH is given by 
\begin{eqnarray}\label{dell}
\Delta L &=& \frac{GM}{6c^2}\left(R_{{\rm Sbrane}}^{-1} - R_S^{-1}\right) \dot{M} c^2\nonumber\\
&=& \frac{1}{12}\left(\frac{R_S}{R_{\rm Sbrane}} -1\right)\dot{M}c^2
\end{eqnarray}\noindent 
For a typical SMBH eq.(\ref{dell}) reads
\begin{equation}
\Delta L \sim  10^{28}\;{\rm erg\, s^{-1}}.
\end{equation}
In terms of solar luminosity units $L_\odot = 3.9 \times 10^{33} {\rm erg\,s^{-1}}$ it follows that
the variation of luminosity of a (SMBH) quasar due to the correction of the Schwarzschild radius in a brane-world scenario is 
given by 
\begin{equation}\label{de1}
\Delta L \sim  10^{-5}\;L_\odot.
\end{equation}

Naturally, this small correction in the Schwarzschild radius of SMBHs given by eqs.(\ref{dell},\ref{de1}) 
implies in a consequent correction in quasar luminosity via accretion mechanism. 
This correction has shown to be a hundred thousand weaker than the solar luminosity. 
In spite of the huge distance between quasars and us, it is probable these 
corrections can be never observed, although they indeed exist in a brane-world scenario. 
This correction is clearly regarded in the luminosity integrated in all wavelength. 
We look forward the detection of these corrections in particular selected wavelengths, since quasars also use to emit radiation
in the soft/hard X-ray band.

In the graphics below we illustrate the variation of luminosity $\Delta L$ of quasars as a function of the SMBH mass and $\ell$, and 
also for a given BH mass, $\Delta L$ as is depicted as a function of $\ell$. 

\begin{figure}
\includegraphics[width=8.5cm]{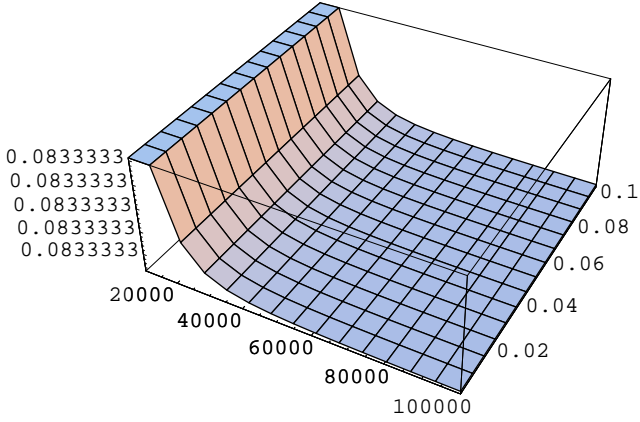}
  \caption{\small 3D graphic of $\frac{\Delta L}{\dot{M}c^2} \times \ell \times M$ where the SMBH mass $M$ varies from 10 to 10$^6$ $M_\odot$
and the radius $\ell$ of the AdS$_5$ bulk varies from $10^{-7}$ to $10^{-1}$ mm.}
\label{fig:ads51}
\end{figure}

\begin{figure}
\includegraphics[width=8.5cm]{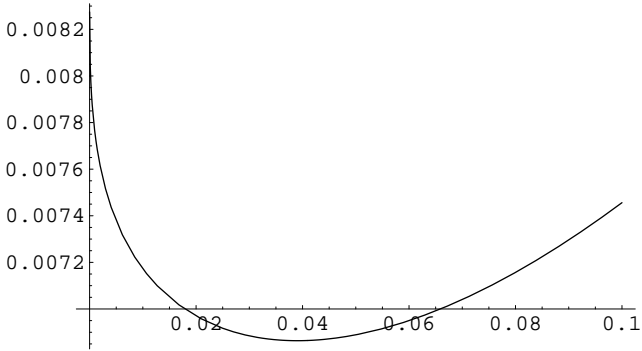}
  \caption{\small  Graphic of $\displaystyle\frac{\Delta L}{\dot{M}c^2} \times \ell$ for $M = M_\odot$}
\label{fig:ads52}
\end{figure}

\begin{figure}
\includegraphics[width=8.5cm]{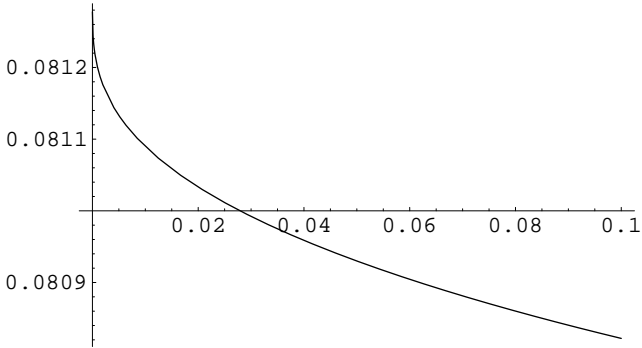}
  \caption{\small Graphic of $\displaystyle\frac{\Delta L}{\dot{M}c^2} \times \ell$ for $M = 10M_\odot$}
\label{fig:ads53}
\end{figure}

\begin{figure}
\includegraphics[width=8.5cm]{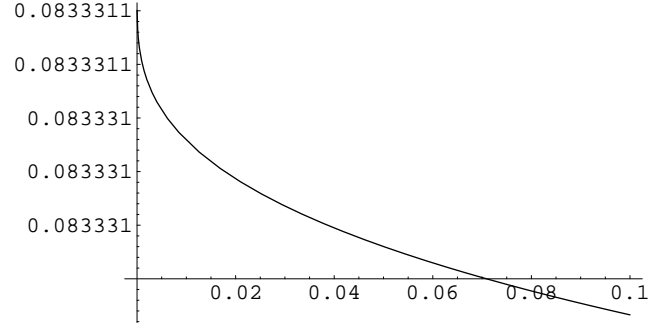}
  \caption{\small Graphic of $\displaystyle\frac{\Delta L}{\dot{M}c^2} \times \ell$ for $M = 100M_\odot$}
\label{fig:ads54}
\end{figure}

\begin{figure}
\includegraphics[width=8.5cm]{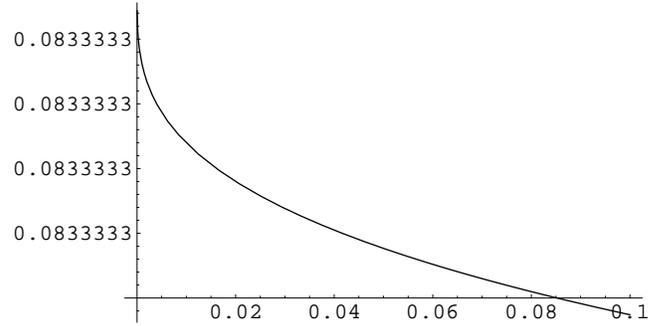}
  \caption{\small Graphic of $\displaystyle\frac{\Delta L}{\dot{M}c^2} \times \ell$ for $M = 1000M_\odot$}
\label{fig:ads54}
\end{figure}

\section{Concluding Remarks and Outlooks}

In the present model the variation of quasar luminosity is regarded as an extra dimension brane effect, and can be immediately estimated
by eq.(\ref{dell}), involving the Schwarzschild radius calculated in a brane-world scenario, and the standard 
Schwarzschild radius of a BH in 3-brane.
It is desirable also to calculate the variation of quasar luminosity in a Kerr and in a Reissner-Nordstr\o m (RN) geometry, where while the latter is caused by a 
electrically neutral, rotating BH, the former is generated by a charged, static black-hole. It shall be done in a sequel paper, using 
a formul\ae\, equivalent to eq.(\ref{h}) but now concerning RN metric \cite{rs06,rs09,Da}.

 The 2-brane model by contrast, for suitable choice of the extra dimension length and of $\ell$ does predict tracks and/or
 signatures in LHC \cite{Maartens,lhc}. Black holes shall be produced in particle collisions at energies possibly below the Planck scale. 
ADD brane-worlds \cite{ad1,ad2,ad3} also provides a possibility to observe black hole production signatures in the 
next-generation colliders and cosmic ray detectors \cite{lhc1}. In the sequel article we will show 
that, since mini-BHs possess a Reissner-Nordstr\"om-like effetive behavior under gravitational potential, they 
 feel a 5$D$ gravity and are more sensitive to extra dimension brane effects.

\section{Acknowledgements}
The authors are grateful to Prof. Paul K. Townsend for his comments and suggestions, to Prof.
Roy Maartens for his patience and clearing up expositions concerning branes.  
The authors thank to CAPES and CNPq for financial support.


\begin{thebibliography}{99}




\bibitem{gr} Green M B, Schwarz J H, and Witten E, \emph{Superstring Theory}, vols. I \& II, Cambridge Univ. Press,
 Cambridge 1987.

\bibitem{zwi} Dienes K R, \emph{String theory and the path to unification: a review of recent developments}, Phys. Rep. {\bf 287}, 447-525 (1997) [{\tt hep-th/9602045}].

\bibitem{zwi1} Kaku M, \emph{Strings, Conformal Fields and M-theory}, Springer-Verlag, New York 2000.

\bibitem{zwi2} Kiritsis E, \emph{Introduction to Superstring Theory}, Leuven Notes in Math. and Theor. Phys. {\bf 9}, Leuven Univ. Press, Leuven 1997 [{\tt hep-ph/9709062}].

\bibitem {Randall1} Randall L and Sundrum R, \emph{An alternative to compactification}, Phys. Rev. Lett. {\bf 83}, 4690-4693 (1999) [{\tt hep-th/9906064}].

\bibitem {Randall2} Randall L and Sundrum R, \emph{A large mass hierarchy from a small extra dimension},  Phys. Rev. Lett. {\bf 83} (1999) 3370-3373 [{\tt hep-ph/9905221}].

\bibitem{rov} Rovelli C, \emph{Loop quantum gravity},  Living Rev. Rel. {\bf 1}, 1-34 (1998) [{\tt gr-qc/9710008}].

\bibitem{Townsend} Townsend P K, \emph{Brane surgery}, Nucl. Phys. {\bf B} Proc. Suppl., {\bf 58}, 163-175 (1997) [{\tt hep-th/9609217}].

\bibitem{ken}  Akama K, \emph{Pregeometry} in Lecture Notes in Physics, 176, \emph{Gauge Theory
and Gravitation},  Kikkawa K, Nakanishi N, and Nariai H (eds.),  Proceedings, Nara 1982
 pp. 267-271 Springer-Verlag, Berlin 1983,  also available in  \emph{`An early proposal of brane
world'} [{\tt hep-th/0001113}].





\bibitem{Likken} Lykken J and Randall L, \emph{The shape of gravity}, \emph{JHEP} {\bf 06}  (2000) 014 [{\tt hep-th/9908076}].

\bibitem{Gregory} Gregory R, Rubakov V A, and Sibiryakov S M, \emph{Brane worlds: the gravity of escaping matter}, Class. Quant. Grav. {\bf 17}, 4437-4449 (2000)
[{\tt hep-th/0003109}].

\bibitem{Maartens} Maartens R, \emph{Brane-world gravity}, Living Rev. Relat. {\bf 7}, 7 (2004) [{\tt gr-qc/0312059}].

\bibitem{ma}  Seahra S S, Clarkson C, and Maartens R, \emph{Detecting extra dimensions with gravity wave spectroscopy: the black string brane-world}, Phys. Rev. Lett. 
{\bf 94} (2005) 121302  [{\tt gr-qc/0408032}].


\bibitem{soda} Kanno S and Soda J, \emph{Black String Perturbations in RS1 Model}, in 
 4th Australasian Conference on General Relativity and Gravitation, Monash University, Melbourne, January 2004.
 To appear in the proceedings, in General Relativity and Gravitation.

\bibitem{emparan} Emparan R, Garcia-Bellido J, and Kaloper N, \emph{Black hole astrophysics in AdS braneworlds}, \emph{JHEP} {\bf 0301} (2003) 079 [{\tt hep-th/0212132}].

\bibitem{Shiromizu} Shiromizu T, Maeda K, and Sasaki M, \emph{The Einstein equations on the 3-brane world}, Phys. Rev. {\bf D62}, 024012 (2000) [{\tt gr-qc/9910076}].

\bibitem{Da} Dadhich N, Maartens R, Papadopoulos P,  and Rezania V, \emph{Black holes on the brane}, Phys. Lett. {\bf B487}, 1-6 (2000) [{\tt hep-th/0003061}].















\bibitem{rs05} Kanti P and Tamvakis K, \emph{Quest for Localized 4-D Black Holes in Brane Worlds}, 
Phys. Rev. {\bf D65}, 084010 (2002) [{\tt hep-th/0110298}].

\bibitem{rs06} Casadio R, Fabbri A, and  Mazzacurati L, \emph{New black holes in the brane-world?}, Phys. Rev. {\bf D65}, 084040 (2002) [{\tt gr-qc/0111072}]. 

\bibitem{rs07} Kanti P, Olasagasti I, and Tamvakis K, \emph{Quest for Localized 4-D Black Holes in Brane Worlds. II : Removing the bulk singularities}, 
Phys. Rev. {\bf D68}, 124001 (2003) [{\tt hep-th/0307201}]. 

\bibitem{rs08} Shiromizu T and Shibata M, \emph{Black holes in the brane world:Time symmetric initial data}, 
 Phys. Rev. {\bf D62}, 127502 (2000) [{\tt hep-th/0007203}]; 
 Chamblin A, Reall H S, Shinkai H A, and Shiromizu T, \emph{Charged brane-World black holes}, Phys. Rev. {\bf D63}, 064015 (2001) [{\tt hep-th/0008177}]. 

\bibitem{rs09} Casadio R and Mazzacurati L, \emph{Bulk shape of brane-world black holes}, Mod. Phys. Lett. {\bf A18}, 651 (2003) [{\tt gr-qc/0205129}]. 

\bibitem{rs01} Germani C and  Maartens R, \emph{Stars in the braneworld}, Phys. Rev. {\bf D64} 124010 (2001) [{\tt hep-th/0107011}].

\bibitem{rs02} Deruelle N, \emph{Stars on branes: the view from the brane} [{\tt gr-qc/0111065}].

\bibitem{rs03} Visser M and Wiltshire D L, \emph{On-brane data for braneworld stars}, Phys. Rev. {\bf D67}, 104004 (2003) [{\tt hep-th/0212333}].


\bibitem{Gian} Giannakis I and Ren H, \emph{Possible extensions of the 4D Schwarzschild horizon in the 5D
brane world}, Phys. Rev. {\bf D63}, 125017 (2001) [{\tt hep-th/0010183}].


\bibitem{card} Fujii K, \emph{A Modern Introduction to Cardano and Ferrari Formulas in the Algebraic Equations} [{\tt quant-ph/0311102}]. 

\bibitem{rcp} Coimbra-Ara\'ujo C H, da Rocha R and Pedron I T, \emph{Anti-de Sitter curvature radius constrained by quasars in brane-world scenarios}, 
Int. J. Mod. Phys. {\bf D14}, to appear [{\tt astro-ph/0505132}]. 

\bibitem{Bondi} Font J A and   Ib\'a\~nez J M, \emph{Non-axisymmetric relativistic Bondi-Hoyle accretion on to a Schwarzschild black hole},
Monthly Not. Royal Astron. Soc., {\bf 298} (3) 835-846 (1998) [{\tt astro-ph/9810344}].

\bibitem{shapiro} Shapiro S L and Teukolski S A, \emph{Black holes, white dwarfs, and neutron stars},  Wiley-Interscience, New York 1983.

\bibitem{lhc}Cavaglia M, Das S, and Maartens R, \emph{Will we observe black holes at LHCs?}, 
Class. Quant. Grav. {\bf 20} (2003) L205-L212.

\bibitem{ad1} Arkani-Hamed N, Dimopoulos S, and Dvali G, \emph{The hierarchy problem and new dimensions at a millimeter}, 
Phys. Lett. {\bf B429}, 263-272 (1998) [{\tt hep-ph/9803315}].

\bibitem{ad2} Antoniadis I, Arkani-Hamed N, Dimopoulos S, and Dvali G, \emph{New dimensions at a millimeter to a Fermi
 and superstrings at a TeV}, Phys. Lett. {\bf B436}, 257-263 (1998) [{\tt hep-ph/9804398}].

\bibitem{ad3} Antoniadis I, \emph{A possible new dimension at a few TeV}, Phys. Lett. {\bf B246}, 377-384 (1990).

\bibitem{lhc1} Cavaglia M, \emph{Black hole and brane production in TeV gravity: A review}, 
Int. J. Mod. Phys. {\bf A18}, 1843-1882 (2003) [{\tt hep-ph/0210296}].



\end{thebibliography}
\end{document}